\documentclass[aps,pra,reprint]{revtex4-1}
\usepackage{graphicx}
\usepackage{amsmath}
\usepackage[utf8]{inputenc}
\usepackage{xcolor}
\usepackage{physics}
\usepackage{overpic}
\usepackage{hyperref}

\def\out{{\color{blue}\fbox{$\nearrow$}}}
\def\link#1{\href{#1}{\out}}
\usepackage{color} 
\definecolor{gray}{rgb}{0.5,0.5,0.5}

\def\e#1{{\rm e}^{#1}}
\def\i{{\rm i}}
\def\d{{\rm d}}
\def\eps{\varepsilon}

\def\delA{\delta\hspace{-0.03em}A}
\def\delt{\delta t}

\def\Cavedef{\overline{C}}
\def\Cave{\overline{\mathbf{C}}}

\begin{document}
\title{Perspectives for analyzing non-linear photo ionization spectra with deep neural networks trained with synthetic Hamilton matrices}
\author{Sajal Kumar Giri, Lazaro Alonso, Ulf Saalmann, and Jan Michael Rost}
\affiliation{Max Planck Institute for the Physics of Complex Systems, N\"othnitzer Stra\ss e 38, D-01187 Dresden, Germany}

\begin{abstract}
We have constructed deep neural networks, which can map fluctuating photo-electron spectra obtained from noisy pulses to spectra from noise-free pulses. 
The network is trained on spectra from noisy pulses in combination with random Hamilton matrices, representing systems which could exist but do not necessarily exist. In [Giri\,et\,al., Phys.\,Rev.\,Lett.\ \textbf{124}\,(2020)\,113201] we performed a purification of fluctuating spectra, that is mapping them to those from Fourier-limited Gaussian pulses. 
Here, we investigate the performance of such neural-network-based maps for predicting spectra of double pulses, pulses with a chirp and even partially-coherent pulses pulses from fluctuating spectra generated by noisy pulses. Secondly, we demonstrate that along with a purification of a fluctuating double-pulse spectrum, one can estimate the time-delay of the underlying double pulse, an attractive feature for single-shot spectra from SASE FELs.
We demonstrate our approach with resonant two-photon ionization, a non-linear process, sensitive to details of the laser pulse.
\end{abstract}

\maketitle

\section{Introduction}
Machine learning (ML) has recently been applied not only in physics \cite{dubr18, mebu+19, caci+19}, but more specifically also in strong-field physics \cite{sebr+08,sami+17, gisa+20}. 
 One of the most abundant topic has been the reconstruction of the temporal shape of an ultrashort laser pulse, aided by ML techniques \cite{whch19,zhwh+20,zidi+20}. The most popular technique for this reconstruction have been different variants of streaking techniques which require normally considerable additional experimental effort, namely a Terahertz laser light source. With its help one can generate a large amount of data\,---\,the streaking traces\,---\,which can be processed with ML to extract the attosecond pulse shape \cite{whch19,zhwh+20}. However, also a direct method from single-shot spectra has been introduced \cite{zidi+20}.

In a different vein, a trained neural network has been proposed to represent a (semi-)classical path integral 
for strong-field physics\cite{lizh+20}, replacing the need to explicitly calculate a large number of classical trajectories to eventually determine the photo-ionization cross section, which is, however, still an approximation as it is constructed semi-classically. To supply training data for a network which can represent the full quantum path integral implies most likely a numerical effort that would be higher than calculating observables directly.

In general, training of a deep neural network needs a very large amount of non-trivial training data. To generate them experimentally requires substantial additional effort (see the streaking example above). 
To obtain such data without serious approximations within theory is often prohibitively expensive as in the second example.

Acknowledging this situation, we have invented another approach: To calculate exactly and explicitly (with the time-dependent Schr\"odinger equation) photo-electron spectra with a large number of pulses and artificial systems, 
 for which the calculation can be done very quickly.
In this way we are able to supply learning data consisting of about $10^{7}$ spectra. 
A network, trained with these synthetic systems, is not only able to purify noisy test spectra, unknown to the network but from the same class of synthetic systems the training was performed with. Also ``real'' spectra can be purified, which could come from experiment, or for this work, from a realistic full calculation with parameters for the helium atom.
Moreover, noise is in the context of machine learning applied to non-linear photo-ionization helpful: Photo-excitation and ionization processes are subject to strict angular-momentum selection rules, thereby limiting the coupling of light to matter. If a light pulse contains noise and operates in a non-linear (at least two-photon absorption) regime it will couple to a much larger part of the electron dynamics of the target. This helps to train the mapping better and enlarges the pool of training spectra naturally.

In general, all trained networks we will present map one type of spectrum into another (desired) one for a photo-ionization scenario of which only a few key elements need to be specified: The target system should have an excited
state around the photon energy $\omega_{*}$ above the ground state and intensities of the light pulse should be such that two-photon processes dominate. It is not necessary to know more about the target system as ideally all target systems accessible by the light as specified are covered by the learning space of the SHM. Therefore, one 
can apply a trained network also to an experimental spectrum from noisy pulses without detailed knowledge of the target system.

Once the design for training such networks with synthetic Hamilton matrices (SHMs) is set up, that is, the spectra
for learning have been computed, it is not difficult to construct other maps with new networks, as the major effort
is to supply the learning data which do not have to be changed, while training new networks is computationally relatively cheap. This allows us to provide several mappings in the following to predict spectra for ideal
double, chirped and even highly structured partially coherent pulses from noisy spectra.
Finally, we will introduce a network based mapping for a typical SASE FEL situation: There, single-shot noisy spectra are recorded which depend on further, not explicitly known parameters, e.g., the geometrical orientation of the sample or the time-delay of double pulses used. Considering the latter situation, we reconstruct from noisy spectra simultaneously the noise-free spectra and the time-delay of the double pulse.
While we cannot do this with the accuracy of the designated algorithms as described in the context of streaking above, we do not need any additional information but the spectrum itself. 

The paper is organized as follows: In Sect.\,\ref{sec:prereq} we give details on the representation of the noisy pulses, explain how to construct the SHMs and describe our fast propagation scheme to solve the electronic Schr\"odinger equation to obtain the photo-ionization spectra.
Section \ref{sec:training} details how the network is trained and set up, including measures how to quantify errors in the reconstruction of spectra and a convenient way to parameterize them.
In Sect.\,\ref{sec:purification} we present the predictions of the photo-ionization spectra for various
pulse forms. Section \ref{sec:simu} discusses the single-shot FEL scenario. The paper ends with conclusions in Sect.\,\ref{sec:conclusions}.

\section{Prerequisites}
\label{sec:prereq}

To determine the photo-ionization dynamics we need two elements, the noisy pulses and an efficient way to describe the electron dynamics. In the end we will specify the process we are interested in, namely two-photon absorption.

\subsection{Pulses}
We distinguish between the ``noisy pulses'' which lead to fluctuating spectra and the ``reference pulses'' for which we want to predict spectra.

There are many different possibilities how to incorporate noise into a signal. We choose the partial-coherence method \cite{pfji+10,mopf+11}. With this method one can create noisy pulses whose average over an ensemble has a well-defined pulse shape. As experimentally demonstrated \cite{mopf+11} these kind of pulses represent pulses from SASE FELs well. In the following, we will use the pulse parameterisation 
\begin{subequations}\label{eq:pcm}\begin{align}
\label{eq:pcm_pulse}
f(t) & =N\,G_{T}(t)F_{\tau}(t)\,,
\\
\label{eq:pcm_gauss}
G_{T}(t) & = \e{{-}2{\ln}2\,t^2\!/T^2},
\hfill
F_{\tau}(t) = {\cal F}^{-1}\!\big[\e{{\rm i}\phi(\omega)}
{\cal F}[ \e{{-}t^2\!/\tau^2}\cos(\omega_{*} t)]\big],
\end{align}\end{subequations}
where 
${\cal F}$ and ${\cal F}^{-1}$ are the Fourier transform and its inverse, and $\omega_{*}$ is the carrier frequency. Noise is introduced through random spectral phases $\phi$, uniformly distributed in the interval ${-}\pi\le \phi\le{+}\pi$. The time scale of the fluctuations is given by the coherence time $\tau$, while the Gaussian $G_{T}(t)$ limits the typical pulse duration to $T$. Otherwise, the pulse duration could grow beyond all limits due to the presence of random spectral phases. A specific (deterministic) noise realization we will label with $\phi_{l}(\omega)$. If not stated otherwise, we use $T=3\,$fs and $\tau = 0.5\,$fs in the following. In order to deal with comparable pulses, we use the normalisation constant $N$ to fix the pulse energy $E_\mathrm{p}$, which would otherwise fluctuate from realisation to realisation.

Any reasonable pulse can serve as a reference pulse, for which the 
map created by the network can predict the spectrum. Reasonable means in the present context that the reference pulse's frequency spectrum is covered by the learning space of fluctuating spectra. The simplest choice is the Gaussian $G_{T}(t)$ in \eqref{eq:pcm} itself rendering the prediction equivalent to removing the fluctuations from the spectrum. Therefore, we call this type of map ``purification'' \cite{gisa+20}. In Sect.\,\ref{sec:simu} we will purify fluctuating spectra from double pulses.
 
\subsection{Paradigmatic 1-dimensional strong-field electron dynamics}

Although the subsequent scheme to construct SHMs is general, for the sake of clarity we will describe it for the processes we will consider as an example, namely two-photon absorption in a helium atom. Thereby, the carrier frequency $\omega_{*}$ of the laser is chosen to be quasi-resonant with the transition energy to the first optically allowed excited state. 

A simple and convenient way to realize this concept is to consider 1-dimensional dynamics with a soft-core potential. The corresponding active one-electron Hamiltonian for helium is given by
\begin{equation}\label{eq:h0}
H_{0} = \frac{\hat p^{2}}{2} + V(x) = \frac{1}{2}\left(\i\frac{\d}{\d x}\right)^{2} -\frac{1}{\sqrt{x^{2}{+}a^{2}}},
\end{equation}
with the soft-core parameter $a=1/\sqrt{2}$ which gives a ground-state energy $E_{0}\,{=}\,{-}$24.2\,eV, close to the ionization potential of real helium (24.6\,eV). We represent the Hamiltonian on a grid
$x_j=j\,\Delta x$, with $\Delta x=0.067$\,a.u. and $x_{\rm max}=500$\,a.u. and determine by diagonalization the eigenenergies $H_{0}|\alpha\rangle=|\alpha\rangle\,\widetilde{E}_{\alpha}$ from the ground state up to $\widetilde{E}_{\alpha}\,\le\,E_{\rm max}\approx 48$\,eV, resulting in 600 eigenstates.

With these eigenstates we calculate the matrix of the time-dependent Hamiltonian $H(t) = H_{0}+ A(t)\hat p$ in velocity gauge
\begin{equation}\label{eq:1Dhammatrix}
\widetilde{H}_{\alpha\beta}(t) = \widetilde{E}_{\alpha}\,\delta_{\alpha\beta}+A(t)\,\widetilde{V}_{\alpha\beta}
\qquad\mbox{with}\quad
\widetilde{V}_{\alpha\beta}\equiv\big\langle\alpha\big|\i{\textstyle\frac{\d}{\d x}}\big|\beta\big\rangle
\end{equation}
with the vector potential $A(t)=A\,f(t)$, $A$ being the field amplitude.

\subsection{Synthetic Hamilton Matrices (SHMs)}
Since we want to train our network such that it recognizes almost arbitrary systems, which only need to have a (quasi-)resonant  transition energy for the first absorbed photon, 
we create SHMs by randomly changing energies ${E}_{\alpha}$ and matrix elements ${V}_{\alpha\beta}$ about the 1-dimensional example system defined in Eqs.\,\eqref{eq:h0} and \eqref{eq:1Dhammatrix} through the variation of four parameters in
\begin{subequations}\label{eq:random}
 \begin{align}
	E_\alpha & = 3^{[\xi_1{-}\gamma]}\widetilde{E}_\alpha &\mbox{for }& \widetilde{E}_\alpha{<}0,\;\alpha{>}0, 
	\\
	V_{0\alpha} & = 3^{\xi_2}\widetilde{V}_{0\alpha} &\mbox{for }& \widetilde{E}_\alpha{<}0,
	\\
	V_{\alpha\beta} & = 3^{\xi_3}\widetilde{V}_{\alpha\beta} &\mbox{for }& \widetilde{E}_\alpha{<}0,\;\widetilde{E}_\beta{>}0,
\\
	V_{\alpha\beta} & = 3^{\xi_4}\widetilde{V}_{\alpha\beta} &\mbox{for }& \widetilde{E}_\alpha{>}0,\;\widetilde{E}_\beta{>}0\,.
 \end{align}
\end{subequations}
Here, $\xi_{i=1\ldots4}=[-1,+1]$ are four uniform random numbers which lead to a large variety of artifical systems with different bound-state energies (\ref{eq:random}a) and couplings between ground and bound states (\ref{eq:random}b), as well as between bound and free states (\ref{eq:random}c) and among free states (\ref{eq:random}d), respectively. 
Finally, with the parameter $\gamma$ the condition of resonant first-photon absorption can be met. In the present case the energy difference between ground and the excited state is equal to the central laser frequency $\omega_{*}$, i.\,e., $E_1\,{-}\, E_0 = \omega_{*}$ if $\gamma = 0.891$ and $\xi_1 = 0$. 
 Note, that $\gamma$ does normally not hamper the application to experimental situations, as one typically knows the binding energy and the central photon frequency. Finally, we construct SHM $H_{\alpha\beta}(t)$ inserting $E_\alpha$, and $V_{\alpha\beta}$ into Eq.\,\ref{eq:1Dhammatrix}.
 
The idea of SHM is an essential part of our approach which serves two purposes: (i) it allows us to supply a sufficient number of theoretical learning data for the network and (ii) it represents a large variety of systems which could exist in nature but not necessarily do so. The SHM should be ``dense enough'' in the parameter space such that always the Hamilton matrix of a real system one is interested in can be interpolated between SHMs, as interpolation capability is a strength of neural networks (in contrast to extrapolation). Of course, one can formulate more sophisticated SHMs with more parameters, but for the present case the four random parameters are sufficient.
 
Yet, we need to overcome one final obstacle, and that is the calculation of the spectra based on the SHMs. To obtain those spectra for arbitrary pulse forms $A(t)$ requires to solve the time-dependent Schr\"odinger equation (TDSE) which in turn implies that we need an extremely fast propagation scheme to be able to solve of the order of $10^{7}$ TDSEs in a reasonable time.

\subsection{Fast solution of the TDSE with SHMs}\label{sec:fast}
To achieve high propagation efficiency we make use of the fact that the Hamilton matrix \eqref{eq:1Dhammatrix} depends explicitly on time only through the vector potential $A = A(t)$. Hence, instead of discretizing the time equidistantly, we discretize the vector potential $A_\mathrm{min}\le A \le A_\mathrm{max}$ in $j_\mathrm{max}$ in steps $A_j=j\,\delA$ with $\delA = (A_\mathrm{max}-A_\mathrm{min})/(j_\mathrm{max}-1)$.

With the time-independent Hamiltonian $H^{j}= H_{0}+A_{j}\,\hat p$ we can construct a short-time propagator which is valid over a time span $\delt_{j}$ short enough such that a fixed $A_{j}$ is a reasonable approximation. Therefore, the unitary short-time propagator can be obtained by direct integration,
\begin{equation}\label{eq:Ushort}
U^{j} = \e{-\i H^{j}\delt_{j}}\,.
\end{equation}
The full propagator $U(t_\mathrm{f},t_\mathrm{i})= \prod_{k}U^{j_{k}}$,
is now simply a concatenation of the short-time propagators over respective time spans $\delt_{k}$ (with $k = 1,\ldots,k_\mathrm{max}$) over which the discretised $A_{j}$ hold,
where $\delt_{1}=t_{1}-t_{i}$ and $\delt_{k_\mathrm{max}}= t_{f}-t_{k_\mathrm{max}-1}$.

To make efficient use of the SHMs, it is imperative that we use the matrix elements from \eqref{eq:random} as they do not require explicit integration over wave functions. Hence, we diagonalise $\langle\alpha|H^{j}|\beta\rangle = E_{\alpha}\,\delta_{\alpha\beta}+j\,\delA\,V_{\alpha\beta}$ in the basis of $H_{0}$ to give its eigenenergies $E^{j}_{\gamma}$ and eigenfunctions $\phi^{j}_{\gamma} = \sum_{\alpha}W^{j}_{\gamma\alpha}\phi_{\alpha}$ leading to the short-time propagator
\begin{equation}\label{eq:Ushortmat}
U_{\alpha\beta}^{j}=\sum_{\gamma}
W^{j}_{\alpha\gamma}
\e{-{\i}\,E_{\gamma}^{j}\,\delt_{j}}W^{j}_{\gamma\beta}
\end{equation}
for fixed vector potential $A_{j}$.

Note, that over the entire pulse $A(t)$ certain $A_{j}$ may occur more than once with different time intervals over which they are valid (if the local derivative $\d A(t)/\d t|_{A_{j}}$ is large, the time interval will be small and vice versa). Therefore it is worthwhile to compute the $U^{j}_{\alpha\beta}$ beforehand and keep them stored. They can be used for all pulses (the fluctuating ones as well as the reference one) for a Hamilton matrix specified by the elements \eqref{eq:random}. Furthermore, we do not calculate the full matrix of the propagator which would involve many
 matrix products. It is sufficient to propagate the vector $|0\rangle$ of the initial state (the ground state of the system) which requires only the computation of matrix-vector products. Only in this way we were able to calculate millions of spectra, necessary to train the network.

\section{Training the network}\label{sec:training}
Through training with fluctuating spectra from the SHMs, the deep neural network encodes the dynamics of two-photon absorption spectra with the central photon frequency $\omega_{*}$ for all target systems covered by the SHMs. 
If the network ``sees'' during training a specific class of spectra much more often than representatives of other classes, it will be biased towards those often found spectra once trained. Hence, we have to fill the learning space of spectra (available for training, validating and testing the network) as homogeneously as possible.
 
\subsection{Generating spectra}\label{sec:spectra}

Synthetic Hamilton matrices  which nearly satisfy the resonance condition, i.\,e.,\ $\xi_1=0$ in Eq.\,\eqref{eq:random}, are particularly sensitive to the pulse shape and therefore generate more structured and diverse spectra through nonlinear processes, here resonant two-photon ionization, than SHMs with $\xi_{1}\ne 0$. To sample the space of input spectra as homogeneously as possible, 50\% of the spectra come from SHMs with $\xi_1\approx 0$ and the other 50\% spectra are from SHMs with uniform $\xi_1$, randomly selected in the range $[-1, +1]$. After training on these spectra the network is not biased for $\xi_1$ around zero but works equally well for all $\xi_1$ in the specified range.

We calculate $n_{\rm mat}=40,000$ reference spectra from the same number of SHMs.
For each reference spectrum, we calculate $n_{\rm pul}\,{=}\,200$ spectra (``fluctuating spectra'') from noisy pulses obtained with the partial-coherence method \cite{pfji+10} using a different noise realization for each SHM. Since solving the TDSE for a single spectrum takes only a few seconds thanks to the highly-optimized propagation scheme outlined in section \ref{sec:fast}, this procedure can be executed despite the need to solve about 10$^{7}$ TDSEs.

For each SHM, we average over all fluctuating spectra $\overline{P}_{k}(E)=\frac{1}{n_{\rm pul}}\sum_lP_{kl}(E)$ instead of using the individual fluctuating spectra $P_{kl}(E)$ computed from $H_{kl}(t)$, where $k$ labels the Hamilton matrix and $l$ the noisy pulse.
We normalize all averaged fluctuating and reference spectra, i.\,e., $\int dE\,P(E)=1$.

The resulting set of $40,000$ averaged fluctuating spectra constitutes a major part of the learning space to train the networks in section 4 for the prediction of spectra from different pulse shapes.

\subsection{Parameterization of spectra and cost functions}
For efficient representation we parameterize each spectrum $\overline{P}_{k}(E)$ in a basis of harmonic oscillator eigenfunctions $\{\chi_{\kappa}\}$,
\begin{equation}\label{eq:spec}
\overline{P}_{k}(E)=\Big|\sum_{\kappa=1}^{n_{\rm bas}}\Cavedef_{k}^{\kappa}\chi_{\kappa{-}1}(E)\Big|^{2}\,,
\end{equation}
with the vector $\Cave\,{\equiv}\,\{\Cavedef_{1}\ldots \Cavedef_{n_{\rm bas}}\}$ of coefficients. A basis size of $n_{\rm bas}\,{=}\,100$ is required for the averaged fluctuating spectra, while for the noise-free spectra $n_{\rm bas}\,{=}\,60$ is sufficient. 

The network maps the coefficients of the fluctuating spectra to those of the predicted underlying noise-free spectrum, 
$\{\Cave_{k}\}\to\{\mathbf{C}_{k}\}$. Goal of the training is to minimize the difference between the predicted vector $\mathbf{C}_{k}$ for the noise-free spectrum and $\mathbf{C}_{k}^{\rm ref}$ of the expected reference spectrum. The coefficients allow us to define a difference familiar from vector spaces as
\begin{subequations}\label{eq:errors}
\begin{equation}\label{eq:distC}
\delta_{\Omega}\equiv\frac{1}{n_{\Omega}}
\sum_{k=1}^{n_{\Omega}}\big[\mathbf{C}_{k}-\mathbf{C}_{k}^{\rm ref}\big]^{2}\,,
\end{equation}
which we use for the cost function in the network training. 
As a measure for the difference of two (normalized) spectra $i$ and $j$ we define their ``distance'' 
\begin{equation}\label{eq:dist1}
	D_{ij}=\int\!\d E\,\big|P_{i}(E)-P_{j}(E)\big|
\end{equation} 
and the average mutual distance
\begin{equation}\label{eq:avdist}
\overline{D}_{\Omega}= \frac{2}{n_{\Omega}(n_{\Omega}{-}1)} \sum_{i>j}D_{ij}
\end{equation} 
within a set of $n_{\Omega}$ spectra.
With
\begin{equation}\label{eq:dist2}
	\eps_{\Omega}\equiv\frac{1}{n_{\Omega}}\sum_{k=1}^{n_{\Omega}}D_{k,k_{\rm ref}},
\end{equation}
\end{subequations}
one can quantify the error in terms of the distance \eqref{eq:dist1} of the spectrum $k$ from the reference spectrum $k_{\rm ref}$, where $\eps \le 2$. The label $\Omega$ stands for the set of data the error is calculated for and can assume the values ``train'', ``val'', or ``test'' for training, validation or test data, respectively.

\subsection{The training setup}\label{sec:train}
\begin{figure}[t]
\centering
\includegraphics[width=.85\columnwidth]{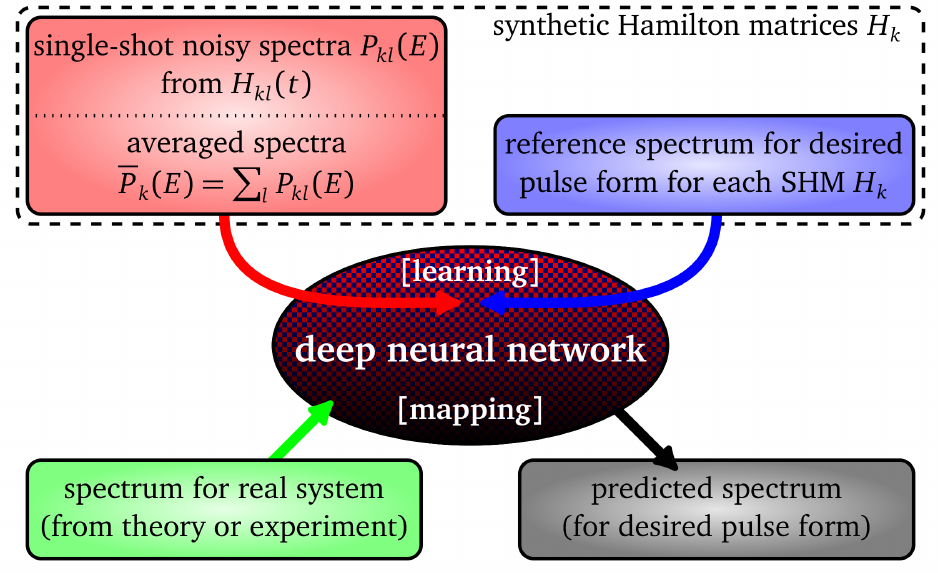}
\caption{Sketch of training and use of a deep neural network with synthetic Hamilton matrices and noisy spectra.}
\label{fig:sketch}
\end{figure}

The full set of learning data contains $n_{\rm mat}\,{=}\,40{,}000$ pairs of spectra. Each pair consists of an averaged noisy spectrum with its respective reference spectrum for the same SHM.
The full learning data set with $n_{\rm mat}$ pairs is split into training (80\,\%), validation (10\,\%) and test (10\,\%) data, respectively. Training corresponds mathematically to minimizing the cost function (\ref{eq:errors}a) with $\Omega\,{=}\,{\rm train}$. Figure \ref{fig:sketch} provides a sketch of what goes into training and prediction.

Implemented with the deep-learning library \textsc{Keras} \cite{ch15}, a fully connected feed-forward artificial neural network is used to establish the mapping. It contains $5$ layers with $60$ neurons on each and was trained at a learning rate of $0.001$ with $100$ epochs, a batch size of $200$ and a learning patience of $25$. 
Each hidden layer neuron contains ReLU activation function\cite{ha19}.
The Adam optimizer \cite{kiba17} is used to minimize the cost function \eqref{eq:distC}. 
The training success is quantified with the error functions \eqref{eq:distC} and  \eqref{eq:dist2}, which both decay logarithmically with the size of the learning data, typical for deep learning \cite{amfu+92, hena+17}.
\begin{figure}[b]
\centering 
\includegraphics[width=\columnwidth]{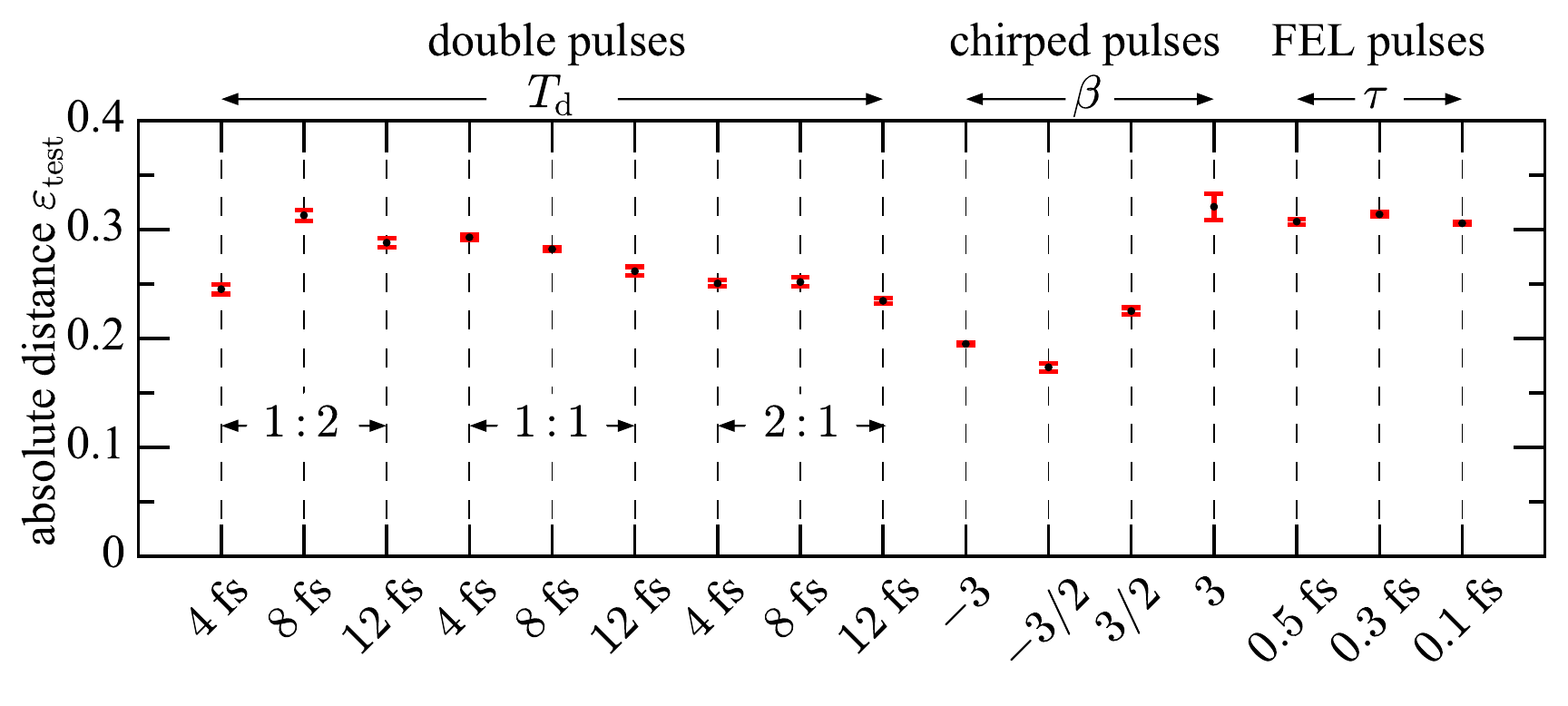}
\caption{Absolute distance error $\varepsilon_{\rm test}$ \eqref{eq:dist2} of the different predicted spectra for test data: \emph{double pulses} with time delay $T_\mathrm{d}$ and amplitude ratios $A_{1}:A_{2}$ as indicated, \emph{chirped pules} with chirp parameter $\beta$, and \emph{partially coherent pulses} with coherence time $\tau$.}
 \label{fig:errors}
\end{figure}%

\begin{figure*}[ht]
\centering
\includegraphics[width=0.34\textwidth]{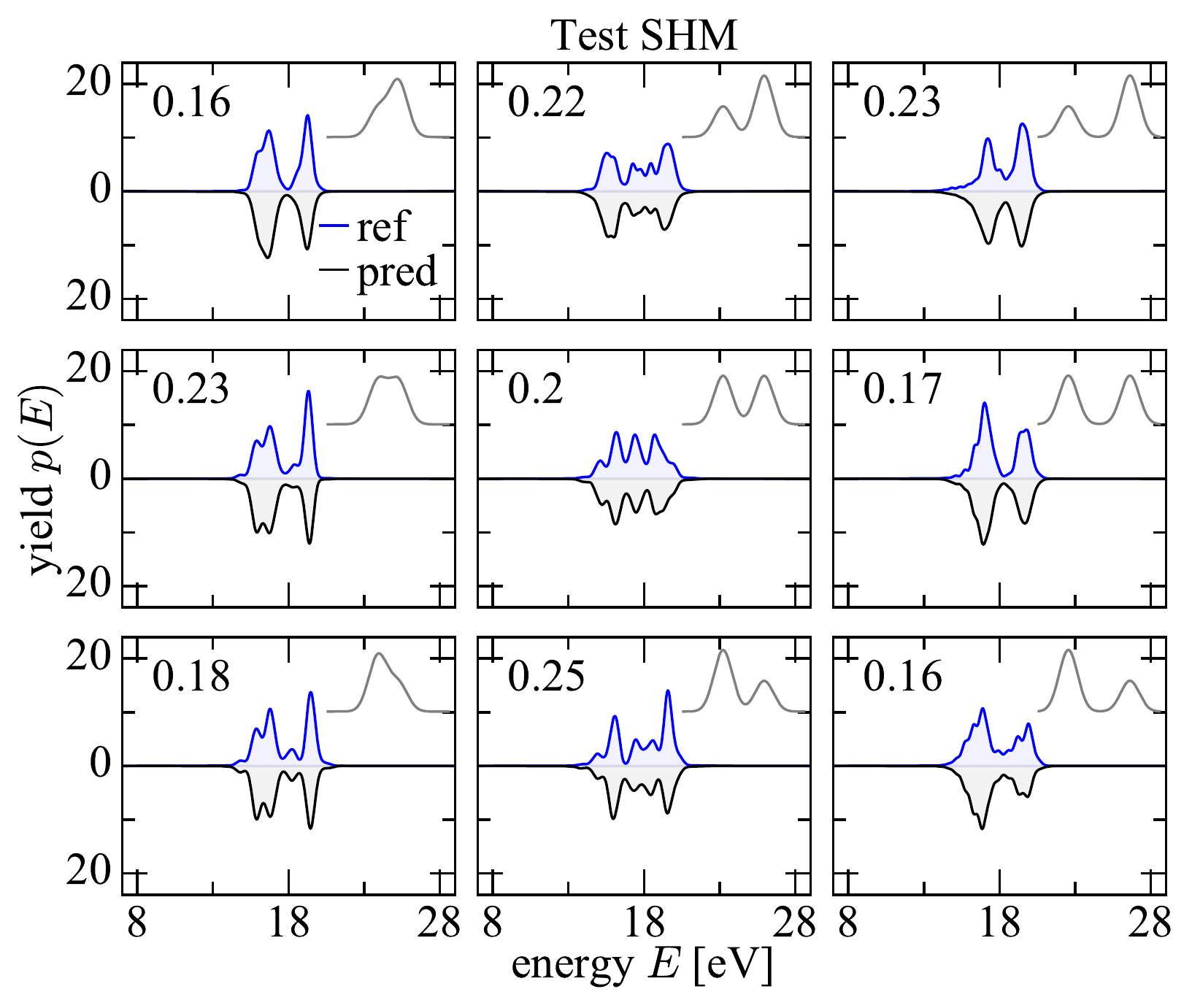}
\hfill\includegraphics[width=0.31\textwidth]{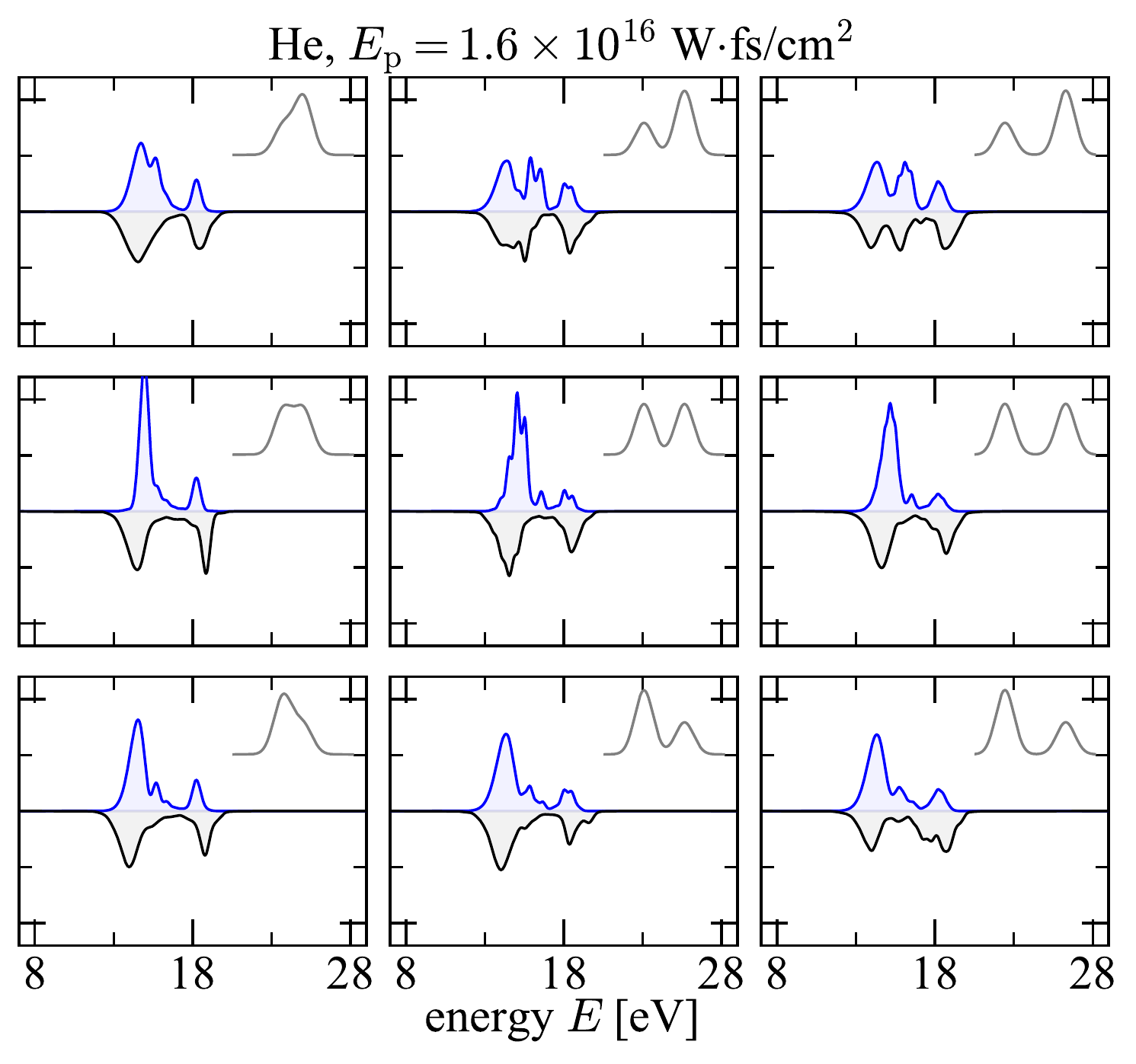} \hfill\includegraphics[width=0.31\textwidth]{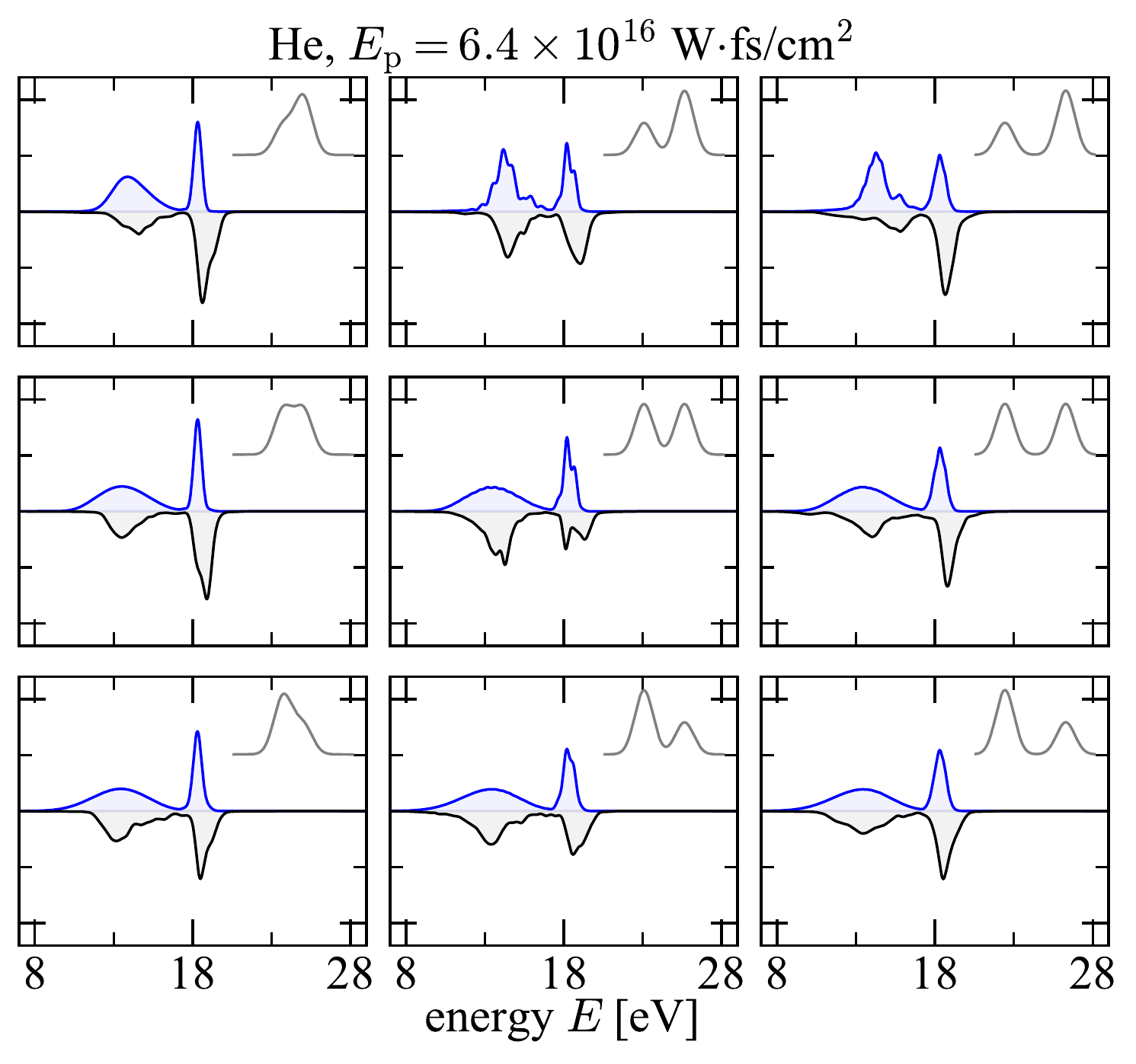}
\caption{Predicted photoelectron spectra (black) are compared to reference spectra (blue). 
All spectra are normalized. The corresponding reference pulses (gray) are shown in each panel. In each of the three figure matrices with 3 x 3 panels, time delays $T_\mathrm{d}$ are 4\,fs, 8\,fs, and 12\,fs, respectively from left to right and from top to bottom, pulse amplitude ratios are $A_{1}:A_{2}$ are 1:2, 1:1, and 2:1, respectively, see \eqref{eq:double}. 
\emph{left matrix:} Prediction for a SHM from test data.
The SHM is chosen such that $E_p\approx 3.84{\times}10^{16}$\,W\,fs/cm$^2$ and each prediction returns an absolute distance (numbers in the panels) $D_{kk_{\rm ref}}$, cf.\,Eq.\,\ref{eq:dist1}, within the range of 30\%\ldots70\% in the error distribution.
\emph{middle matrix:} Prediction of noisy 3D helium spectra (composed of the sum of the two relevant angular-momentum channels s and d) through the trained network for pulses of pulse energy $E_p=1.6{\times}10^{16}$\,W\,fs/cm$^{2}$. 
\emph{right matrix:} Same as middle matrix but for an energy of $E_p=6.4{\times}10^{16}$\,W\,fs/cm$^{2}$.
}
\label{fig:double}
\end{figure*}

\section{Prediction of spectra for different pulse shapes}\label{sec:purification}
To assess the quality of the mapping achieved with the trained networks on the basis of the SHM learning data,
we will discuss scenarios with three different reference pulses for which we predict spectra: (i) double pulses with different time delays $T_\mathrm{d}$ and peak amplitude ratios $A_{1}:A_{2}$, (ii) chirped pulses with chirp parameter $\beta$, and (iii) partially coherent reference pulses with different coherence times $\tau$ according to \eqref{eq:pcm}. 
We have used the network setup for all three scenarios as described in the previous section with the \emph{same} set of fluctuating spectra for training but paired for each SHM with reference spectra which differ corresponding to the above reference pulses. The fluctuating spectra used as input of the network have been generated with the pulses from \eqref{eq:pcm} with a pulse length of $T=3$\,fs, a coherence time of $\tau = 0.5$\,fs, central photon frequency of $\omega_{*}=21$\,eV and 
 intensities between $8\times 10^{15}$\,W\,fs/cm$^{2}$ and $8\times 10^{16}$\,W\,fs/cm$^{2}$.
For further reference and to give an overview how successfully the trained networks can predict spectra for the different pulse shapes from the fluctuating spectra, we show to begin with in Fig.\,\ref{fig:errors} the absolute distance errors ($\epsilon\le 2$) of all predicted spectra. 
Note that for double pulses, the error decreases with increasing time-delay which is probably to be expected since it is easier to identify the time delay if it is larger. The smallest one $T_\mathrm{d}=4\,$fs basically corresponds to a single pulse (recall that the width of each individual pulse is $T=3\,$fs). Interestingly, the sensitivity to the amplitude ratios of the double pulses is even larger than to the time delay: The spectrum from a 1st pulse which is is stronger than the 2nd one is easier to predict than vice versa with pulses of equal strength taking the middle position in terms of the error.

The strongest sensitivity occurs for spectra from chirped pulses where the ones with the most positive chirp ($\beta = {+}3$) are twice as difficult to predict than ones with $\beta = {-}3$. We will come back to this point later. Finally, it is surprising that a spectrum from a partially-coherent pulse, which is naturally very ``busy'', can be identified and therefore predicted from the (averaged) fluctuating spectra, even if the coherence time is shorter than that of the noise ($\tau = 0.5$\,fs) with similar accuracy as for longer coherence times of the reference spectrum. We will discuss the spectra from the different pulse forms now in detail.

\subsection{Prediction of spectra from double pulses}
The reference pulse is here given by 
\begin{equation}\label{eq:double}
f_\mathrm{d}(t)=N_\mathrm{d}\big[A_{1}G_{T}(t+T_\mathrm{d}/2)+A_{2}G_{T}(t-T_\mathrm{d}/2)\big]\cos(\omega_{*}t),
\end{equation}
where $T_\mathrm{d}$ is the delay between the maxima of the two pulses with shape $G_{T}$ from \eqref{eq:pcm_gauss}
and respective amplitudes $A_{i}$. The normalization constant $N_\mathrm{d}$ is used in the same manner as in Eq.\,\eqref{eq:pcm}.

Figure~\ref{fig:double} shows predicted spectra for exemplary double pulses with pulse shapes indicated in gray. Comparison of black and blue curves also helps to develop a sense for what the quantitative distance errors in Fig.\,\ref{fig:errors} mean for the quality of the predictions. The generally good agreement proves that the training of the network was successful and has generated an accurate map.

However, the test data, although not used for training, belong to the same class of SHM used for training. A more realistic test is the prediction of a 3D helium spectrum as shown in Fig.\,\ref{fig:double} (middle), as this is similar to predicting spectra from experimental fluctuating pulses. In general, the prediction works very well, as one can see\,---\,only small details of the spectral structures are sometimes not resolved. This is remarkable, as the shapes of the spectra from the same reference pulses are quite different for the 1D system used for training and the 3D helium (compare the individual equivalent panels of the left and middle Fig.\,\ref{fig:double}). This confirms the transferability of the network and underlines its interpolation capability.

Predictions become worse for increasing pulse energy as shown in the right part of Fig.\,\ref{fig:double}. This is also true for the test data (not shown) but to a slightly lesser extent. While features are still reproduced, the predicted spectra are in general slightly too wide compared to the reference spectra.

\subsection{Prediction of spectra from chirped pulses}
\begin{figure}[ht]
\centering
 \includegraphics[width=0.8\columnwidth]{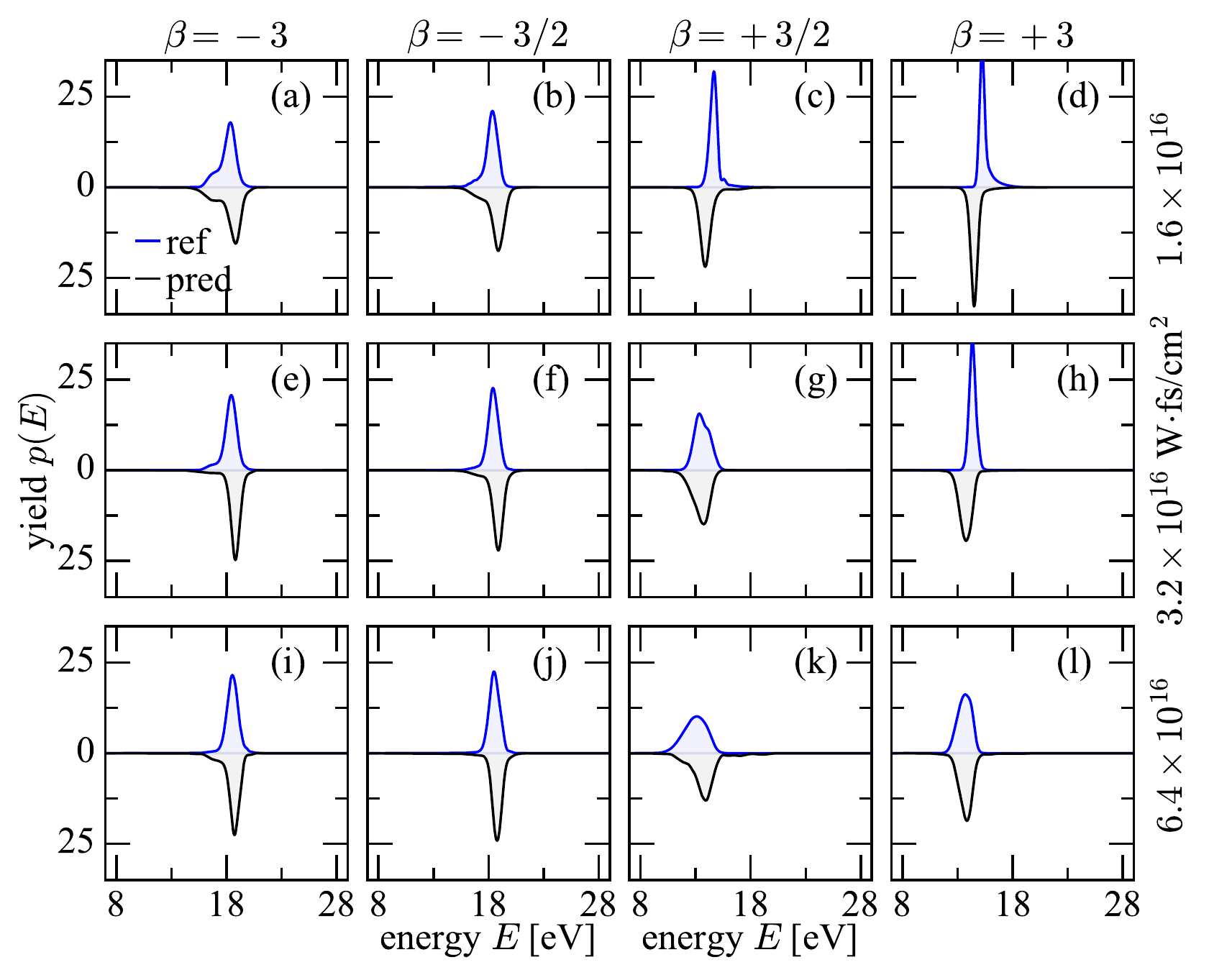}
 \caption{Prediction of 3D helium spectra (black) for chirped pulses \eqref{eq:chirp}. The reference 3D helium spectra are shown in blue.
}
 \label{fig:chirpHe}
\end{figure}

The chirped reference pulses are parameterised by $\beta$ and read 
\begin{subequations}
\label{eq:chirp}
\begin{align}
f_{\beta}(t) & =N_{\beta}G_{\beta}(t)\cos\big(\varphi_{\beta}(t)\big),
\\
\varphi_{\beta}(t) & = \omega_{*}t + \frac{2\ln 2}{\beta + 1/\beta}\frac{t^{2}}{T^{2}},
\quad T_{\beta}=\sqrt{1+\beta^{2}}\,T\,,
\end{align}
\end{subequations}
with the Gaussian from \eqref{eq:pcm_gauss} and $T=3\,$fs. 
Again we normalize the pulse energy, here by means of $N_{\beta}$, as before in Eqs.\,\eqref{eq:pcm} and \eqref{eq:double}.
The predicted spectra are shown in Fig.\,\ref{fig:chirpHe}. They do not exhibit detailed structure, mostly a single peak with different form of the shoulders and reconstruction seems to work well with the exception of large positive chirp, where the position of the spectral peak is systematically red shifted in the predicted spectrum consistent with the largest error (see Fig.\,\ref{fig:errors}) the positively chirped spectra have. 

\subsection{Prediction of spectra from partially-coherent pulses}
\begin{figure}[ht]
\centering
 \includegraphics[width=0.8\columnwidth]{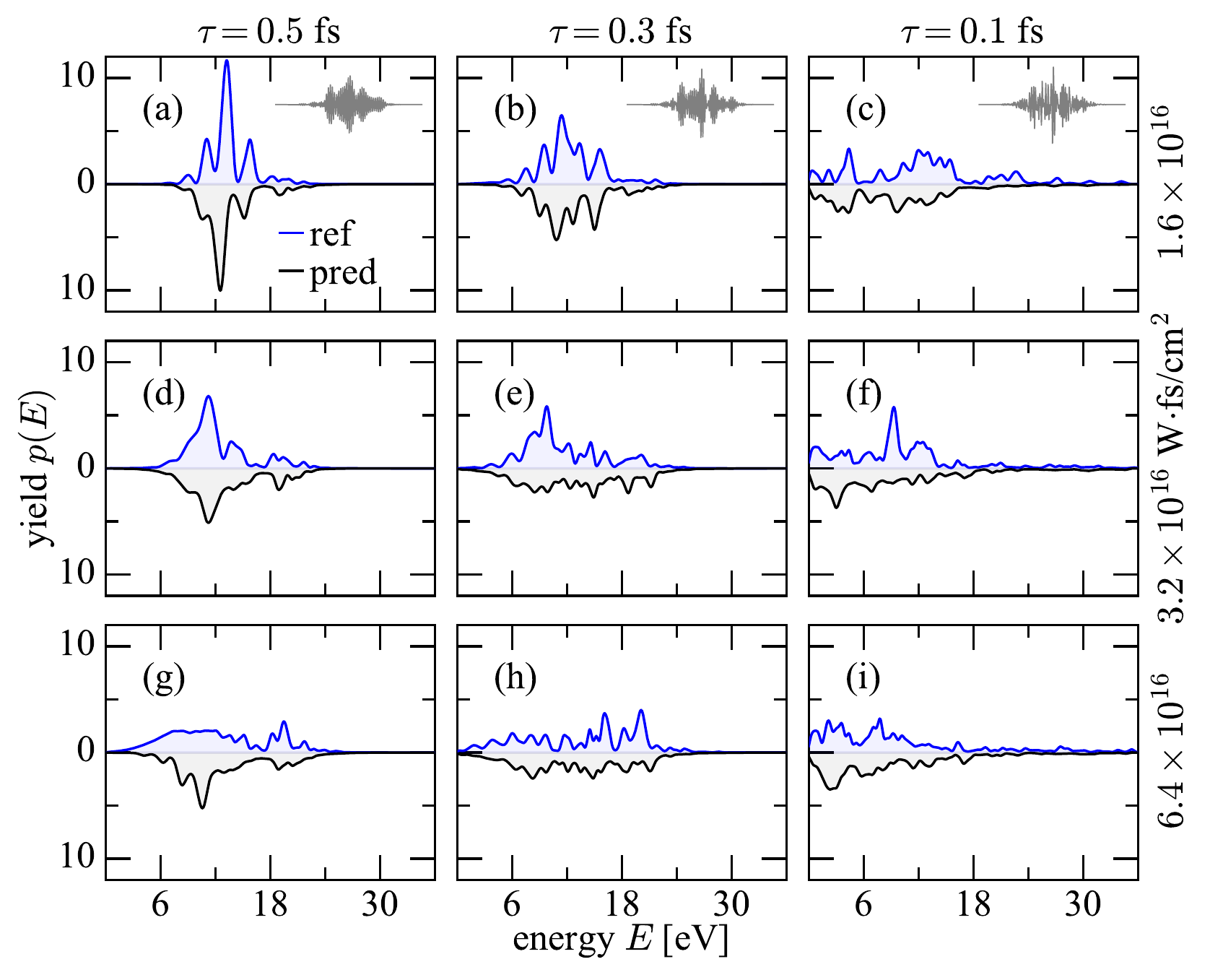}
 \caption{Prediction of 3D helium spectra (black) for partially coherent pulses \eqref{eq:pcm}. The reference 3D helium spectra are shown in blue.
}
 \label{fig:pcsHe}
\end{figure}

We finally will predict spectra from pulses which are themselves ``noisy'', i.e., partially coherent 
and generated according to \eqref{eq:pcm} but for different coherence times $\tau$, typical for SASE FELs. The motivation for such reference spectra was to see where the prediction breaks down since we had the expectation that, at least for spectra from pulses with coherence times much shorter than the ones used for the learning space of fluctuating spectra, the trained network would loose its predictive capability, even more so as the spectra have quite detailed features, see Fig.\,\ref{fig:pcsHe}. However, to our surprise this is not the case, as also revealed by the errors given in Fig.\,\ref{fig:errors}.

\subsection{Prediction errors for different pulse shapes}
\begin{figure}[t]
\centering 
\includegraphics[width=\columnwidth]{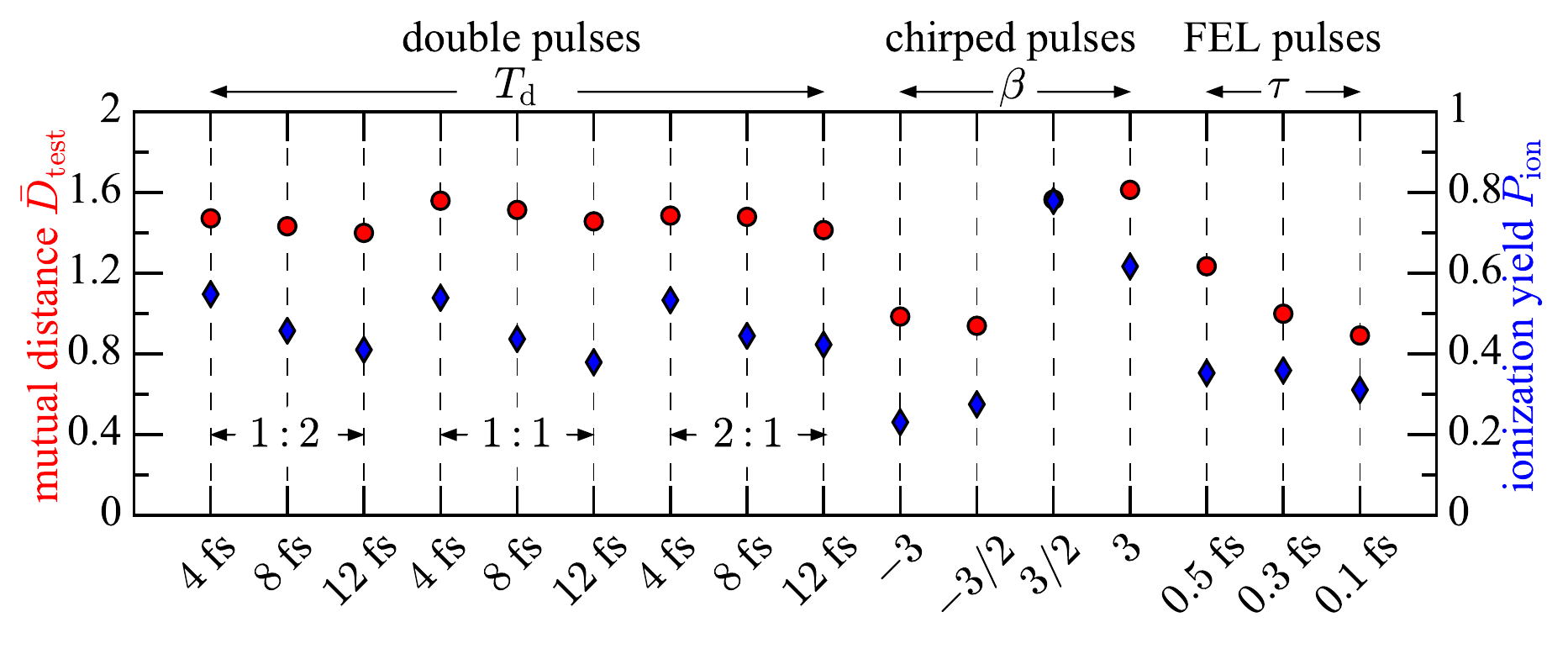}
\caption{Properties of test reference spectra for the pulses from Fig.\,\ref{fig:errors}. Average ionization yield $P_{\rm ion}$ (blue, right axis) and average mutual distance $\overline{D}_{\rm test}$
(red, left axis). 
}
 \label{fig:intens}
\end{figure}

Now, we are in the position to understand details of the distance errors $\eps_{\rm test}$ in Fig.\,\ref{fig:errors} for reference spectra from different pulse shapes. As one can see from Fig.\,\ref{fig:intens}
as a rule of thumb, the smaller the ionization probability $P_{\rm ion}$, the smaller is the diversity of spectra the pulses generate including reference spectra.
 All spectra in this section have been analyzed with networks trained with a learning data set of the same size and a common set of input averaged fluctuating spectra. Therefore, one would expect that the average mutual distance $\overline{D}_{\Omega}$, defined in Eq.\,\eqref{eq:avdist}, of reference spectra is larger for a more extended space of highly diverse spectra as compared to a smaller space of less diverse spectra.
This is indeed the case as $\overline{D}_{\rm test}$ shown with red points in Fig.\,\ref{fig:intens} reveal: They follow the trend of $P_{\rm ion}$ for the test data. Since it is more difficult for the network to interpolate if the available
reference spectra are more distant, one would expect larger errors which explains the trend of the distance errors
in Fig.\,\ref{fig:errors}. In particular striking is the change for chirped pulses\cite{sagi+18}: negative chirp produces small $P_{\rm ion}$
and in turn a moderate diversity of spectra with relatively small $\overline{D}_{\rm test}$ and therefore also the smallest $\epsilon_{\rm test}$.
For positive chirp, the exact opposite holds. One cannot expect that ionization yield, distance of spectra and errors
are directly proportional, as the physical process leading from the pulses to the spectra is still non-linear. For instance, long time-delays in double pulses
give rise to more diverse spectra than short time-delays. Moreover, the $\epsilon_{\rm test}$ are for predictions from noisy spectra. Yet, the causal chain of $ P_{\rm ion}\to \overline{D}_{\rm test} \to \eps_{\rm test}$ holds.

This section has shown that the trained networks can predict spectra from widely varying pulse forms well. The effort one has to invest into the deep neural networks for the prediction of the spectra depends on the diversity
of spectra a certain pulse form is capable to generate.

\section{Single-shot noisy double pulses: 
Simultaneous purification of spectra and reconstruction of time-delay}\label{sec:simu}
The analysis of the previous section has prepared us for the final goal of this work, namely purifying the spectra while simultaneously extracting the correct time-delay from spectra recorded with noisy double pulses which have an unknown time-delay within a certain interval. This scenario is motivated by SASE XFEL pulses\cite{mara+15}, where the pulse is split by a chicane for the relativistic electron bunch which creates the light pulse, or by situations where an XFEL pulse and a time-delayed strong laser pulse are used together whereby the delay between the two pulses is characterized by a jitter from shot to shot. 
 
We model fluctuating double pulses with noise-free double pulses and admixture of noisy double pulses,
\begin{equation}\label{eq:pcm_double}
f_\mathrm{dq}(t)=N_\mathrm{dq}[G_{T}(t{+}T_\mathrm{d}/2)+G_{T}(t{-}T_\mathrm{d}/2)][\cos(\omega_{*}t)+q F_{\tau}(t)],
\end{equation}
where $q = 0.32$, $\tau=0.3$\,fs, $G_{T}$ and $F_{\tau}$ are from \eqref{eq:pcm} and the time-delays $T_\mathrm{d}$ vary between 2\,fs and 14\,fs. Hence, for this task we have to create a new learning space
of fluctuating spectra as input for the network based on fluctuating double pulses. And again, the normalization factor $N_\mathrm{dq}$ ensures the required pulse energy.

Since so far we have not extracted the time-delay of the pulses from the spectra, we verify in section \ref{sec:timedel}, that it is possible to identify the time-delay of double pulses from noise-free spectra generated by those pulses.
In section \ref{sec:singleshot} we will address fluctuating spectra. We first determine the pulses' time-delay $T_{\rm d}$ encoded in
 single-shot spectra generated with noisy double pulses. Subsequently, we average the single-shot spectra with identified $T_{\rm d}$ over small intervals of time-delay (1fs) and purify these averaged spectra. Recall, that purifying means that we remove the fluctuations from spectra by predicting the spectra generated from the respective noise-free pulse forms, in the present case from the noise-free double pulses.

\subsection{Extraction of time-delay from spectra generated with double pulses}\label{sec:timedel}
 
 Here, we aim at constructing a network-based map to extract the time-delays $T_\mathrm{d}$ of double pulses from the (noise-free) spectra the pulses $f_\mathrm{d0}$ from \eqref{eq:pcm_double} generate. To this end we have generated a learning data set of spectra from 20,000 SHMs, each paired with a single double pulse $f_\mathrm{d0}(t)$ with delays between 2 and 14 fs. The learning data is distributed into training, validation and test data as before (see Sect.\,\ref{sec:train}), and the network is also that of Sect.\,\ref{sec:train}, but the number of neurons on each layer is 50, the learning rate is 0.008 and the number of epochs is 200.
 
\begin{figure}[ht]
\centering
 \includegraphics[width=.8\columnwidth]{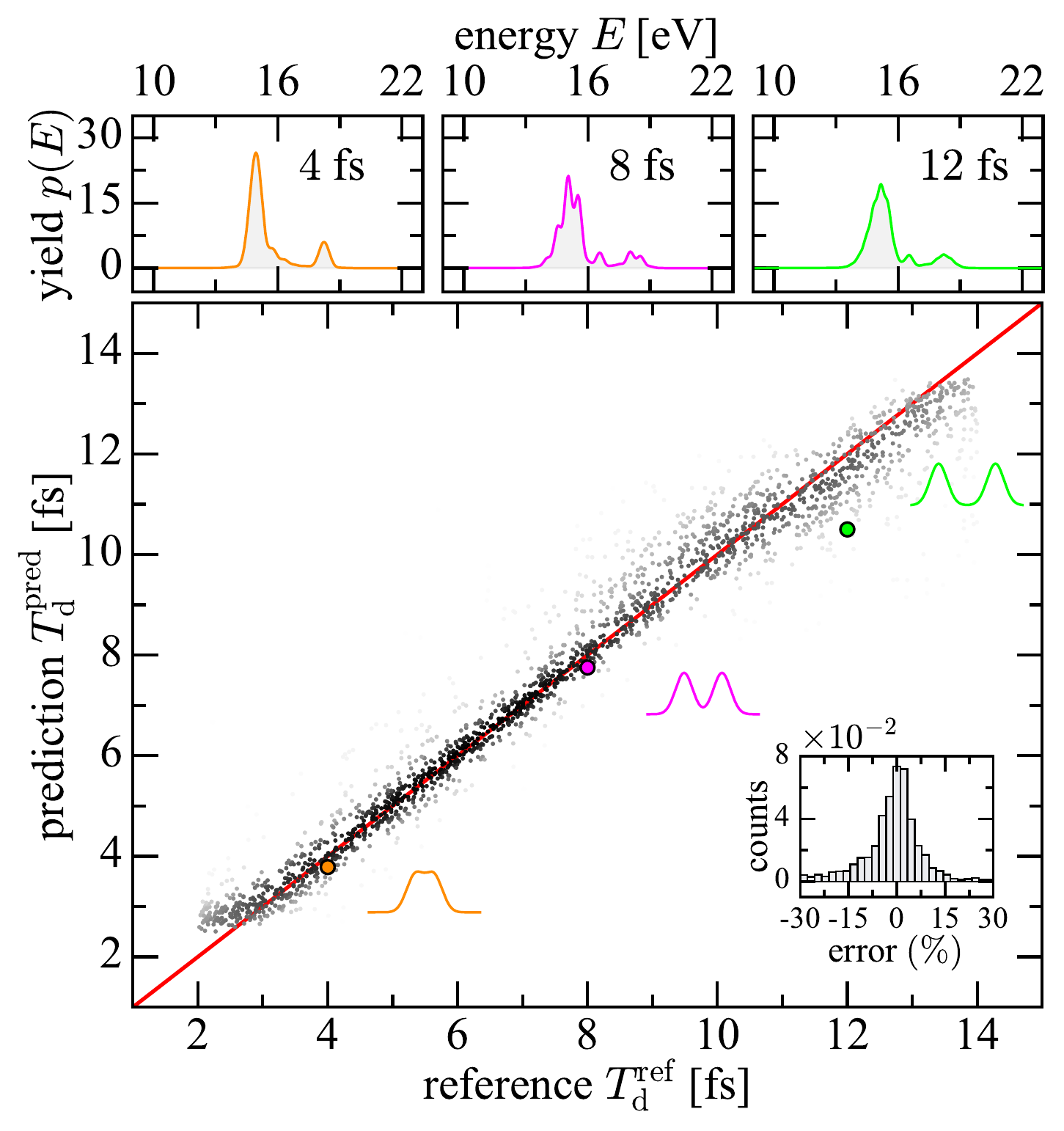}
 \caption{Predicted time-delays against reference time-delays for the test data. 
The pulse energy is $E_\mathrm{p} = 1.6{\times}10^{16}$\,W\,fs/cm$^{2}$. 
 The error distribution of the time-delays for the test data is shown in the lower inset. The red line represents error free prediction. The trained network is transferred to the 3D helium spectra for three time delays: 4\,fs, 8\,fs, and 12\,fs with the reconstructed time delays shown as circles and double pulse shapes sketched. The upper inset gives the corresponding photoelectron spectra. 
}
 \label{fig:delay}
\end{figure}
 
Figure~\ref{fig:delay} shows the training success with the SHMs as well as the transfer of the network to unknown 3D helium spectra. The trained network reproduces well the delays (results scatter along the ideal red line with an error give in the inset). For short $T_\mathrm{d}$ the results deviate from the ideal line since the individual pulses in the double pulse have a width of $T=3\,$fs which limits the resolution towards small time-delays. Results for the reconstructed time-delay for full 3D helium spectra are given for $T_\mathrm{d}$ of 4,\,8, and 12\,fs, respectively, and demonstrate the transferability of the network. The upper row shows the corresponding 3D helium spectra. Given the similarity of these spectra for different time-delays it is remarkable that the trained network can reliably extract the time-delays.
We may conclude that we can map out the delay of the pulse from the spectrum it has produced with the help of the trained network.

\subsection{Purification of single-shot spectra and simultaneous extraction of the time-delay of the generating double pulse}\label{sec:singleshot}

\begin{figure}[ht]
\centering
 {
 \includegraphics[width=.8\columnwidth]{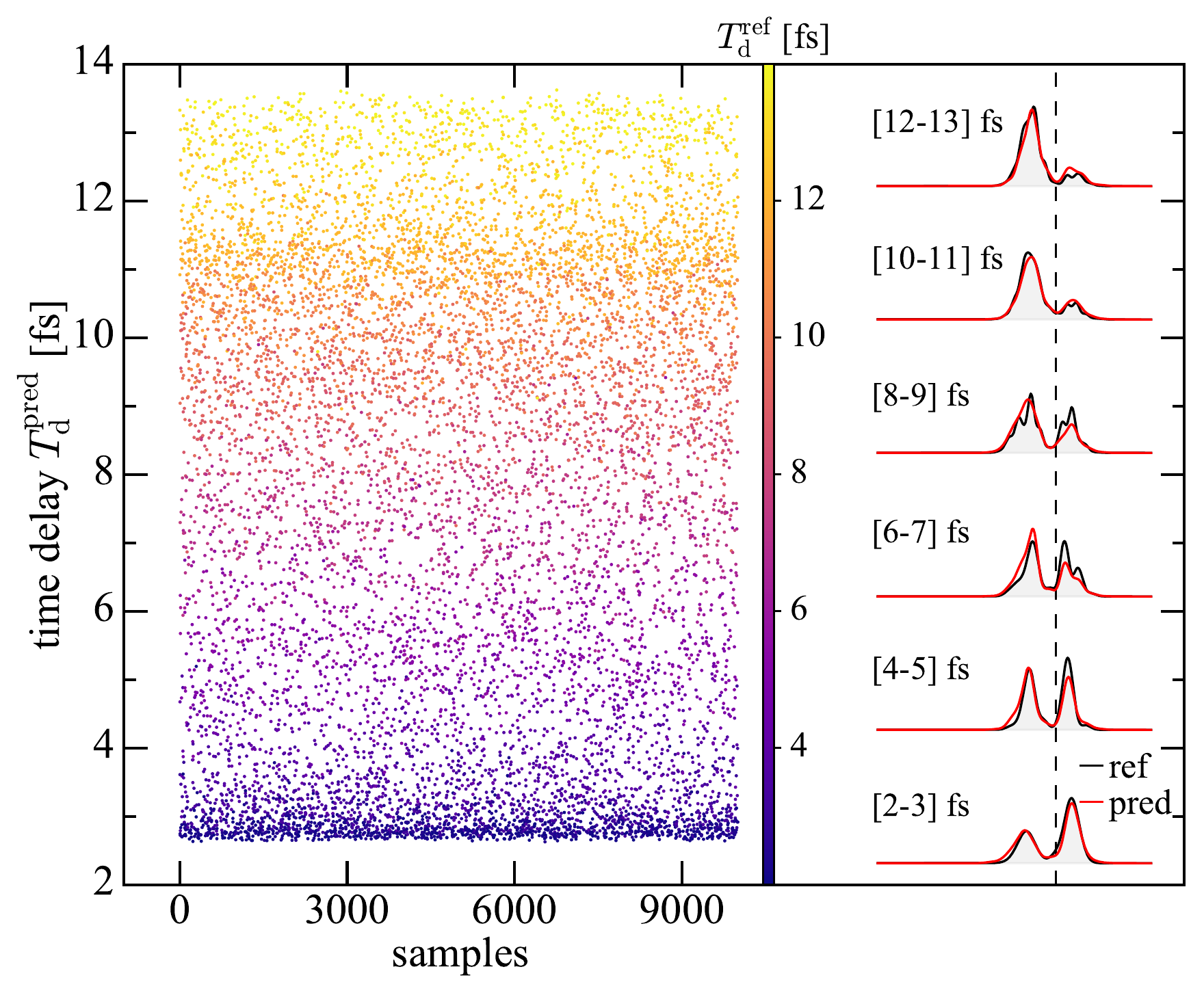}
 }
 \caption{Simultaneous reconstruction of time-delay and purification of noisy spectra for a single Hamilton matrix taken from test data. Single-shot fluctuating spectra for random time-delays are passed through a network to reconstruct the underlying time delays which are shown as scattered points where the color represents the reference time-delay. We consider 12 intervals of time delay in the range 2--14\,fs with interval length of 1\,fs. All single-shot spectra which fall into on interval of time-delay are averaged. The averaged spectra are passed through an another network which maps averaged noisy spectra to purified ones. The predicted purified spectra (red) are compared to reference spectra (black).}
 \label{fig:singleshot}
\end{figure}

Finally, we analyze noisy single-shot spectra with the goal to purify them as in Sect.\,\ref{sec:purification} and to extract the time delay of the generating double pulse as in Sect.\,\ref{sec:timedel} simultaneously. In order to have a reasonable statistics for the map and also for having reasonably different spectra for different time delays, we reconstruct from each noisy single-shot spectrum (all for the same SHM) the time-delay but average the spectra afterwards over small intervals (1\,fs) of time-delays. Subsequently, the averaged spectra are passed through another trained network to purify them. The result is shown in Fig.\,\ref{fig:singleshot}. The scattered points are reconstructed time-delays coloured with the reference time-delays. The monotonous change in colors demonstrates that the reconstruction of time-delays for the test data has been successful. The spectra within 1\,fs intervals of reconstructed time-delays are averaged and subsequently purified. They are shown on the right in red along with reference spectra (black), averaged over the same interval of time-delays. The generally good agreement demonstrates that reconstruction of time-delays and purification of the single-shot spectra is possible without additional information but the single-shot spectra.

\begin{figure}[ht]
\centering
\includegraphics[width=.8\columnwidth]{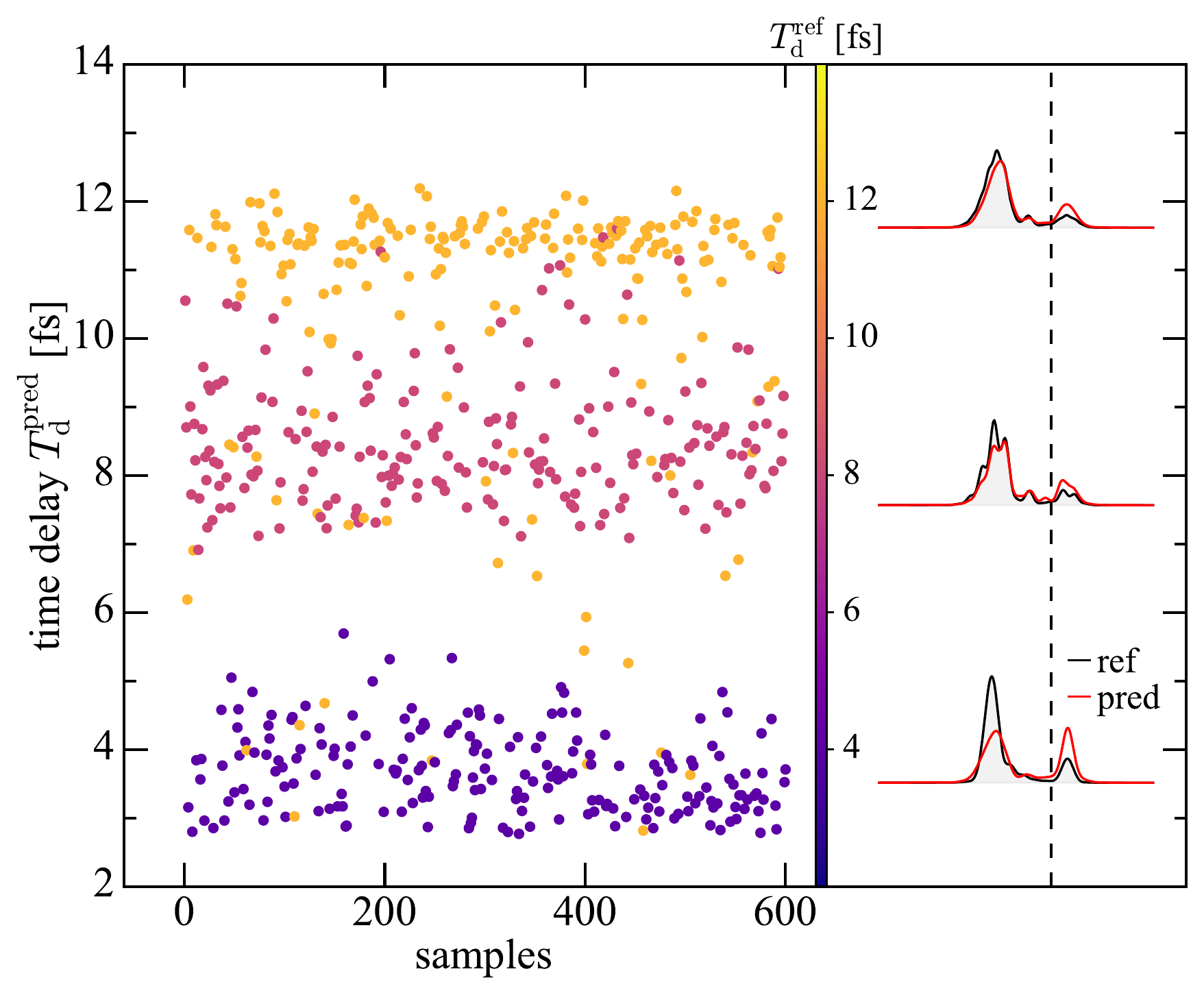}
\caption{Same as Fig.\,\ref{fig:singleshot} but for 3D helium for which the network was not trained. The distribution of predicted time-delays shows three main peaks at 4,\,8,\,12\,fs. The single-shot spectra are averaged over all spectra with time-delays in an interval of 1\,fs about the three peaks. The averaged spectra are 
	passed through the trained network to obtain the corresponding purified spectra shown on the right (red). The three averaged reference spectra (black) are obtained in the same way.}
	\label{fig:hesim}
\end{figure}

The last step is to proof that the reconstruction and purification can be transferred to spectra unknown to the networks. To this end we take noisy single-shot spectra of 3D helium with three well-defined time-delays and pass them through the trained network for reconstruction of the time-delay. The scattered points in Fig.\,\ref{fig:hesim} show the reconstructed time-delays. We average the corresponding spectra over 1\,fs about the three peak time-delays in the scattered points and pass the averaged spectra through the purification network to arrive at the three spectra on the right in red. They agree well with the corresponding reference spectra, averaged over the same intervals of time-delay (black). Hence, the trained networks should be able to reconstruct the time-delay and purify the corresponding fluctuating experimental spectra as they are produced by SASE FELs.

\section{Conclusions}\label{sec:conclusions}
To summarize, we have devised a strategy to create maps through deep neural networks between fluctuating nonlinear photo-ionization spectra and noise-free spectra and between fluctuating single-shot spectra and pulse properties. A crucial part of this strategy is the formulation of synthetic Hamilton matrices which describe artificial systems, similar to ones existing in reality. We use the SHM to generate a sufficient amount of spectra for training the network. In a first application \cite{gisa+20} we purified fluctuating spectra as typically produced by SASE FELs through a neural-network-based map. 

Here, we have taken this mapping capability to a new level by predicting from fluctuating spectra\,---\,which should come ultimately from experiment\,---\,the spectra which would be obtained with specific noise-free pulses, namely double pulses, chirped pulses and chaotic (partially-coherent) pulses. While generally the prediction works as well as the purification (prediction) for simple Gaussian pulses before, the error analysis has revealed interesting differences for the different pulse shapes. 

In a second application we have constructed a neural-network-based map which can extract the time-delay of double pulses from fluctuating single-shot spectra generated by those noisy double pulses. Finally, we could demonstrate that suitably trained networks can achieve both, purification and extraction of the time-delay, from fluctuating single-shot spectra as typically produced by SASE FELs. Clearly, neural networks open promising new ways to analyze in particular noisy data with a potential which has been by far not exhausted.

\section*{Acknowledgements}
We acknowledge financial support by ``BiGmax'', the Max Planck Society's research network on big-data-driven materials science,
and by ``QUTIF'', a priority program (no.\,1840) of the Deutsche Forschungsgemeinschaft.






\providecommand*{\mcitethebibliography}{\thebibliography}
\csname @ifundefined\endcsname{endmcitethebibliography}
{\let\endmcitethebibliography\endthebibliography}{}

\end{document}